# NONLINEAR EXCITATION OF SECOND SOUND IN QUANTUM SOLUTIONS $\text{He}^3 - \text{He}^4$ DUE TO LIGHT WAVE ABSORPTION

## N.I. Pushkina

Moscow State University
Scientific Research Computation Centre
Vorob'yovy Gory, Moscow, 119992 Russia
E-mail: N.Pushkina@mererand.com

*Nonlinear excitation of second sound in superfluid solutions $\text{He}^3-\text{He}^4$ by light waves due to their absorption is discussed. Nonlinear wave equations that model the interaction of second-sound waves with light waves are derived, the expression for the nonlinear interaction length is obtained and an order-of-magnitude numerical estimate of the distance at which a second-sound wave could be amplified from a fluctuation level up to observable values is performed.*

Nonlinear wave interactions that are essentially conditioned by wave absorption in a medium were first investigated in the frame of nonlinear optics [1,2]. For the case of hydrodynamics this type of nonlinear processes was studies in Refs. [3,4]. In the present paper the following absorption-induced nonlinear interaction is considered: two light waves $E_1$ and $E_2$ with slightly different frequencies propagate in a weak superfluid solution $\text{He}^3-\text{He}^4$ at a small angle to each other and due to their absorption a second-sound wave with a frequency equal to the difference frequency of the light waves is excited and amplified.

The Hamiltonian approach will be used for describing second-sound wave propagation in quantum solutions. For superfluid solutions $\text{He}^3-\text{He}^4$ Hamiltonian variables are three pairs of canonically conjugated variables $(\rho, \alpha)$, $(S, \beta)$, $(N, \xi)$ [5]. The meaning of these variables is as follows: $\rho$ is the solution density; the quantity $\alpha$ determines the superfluid velocity $\mathbf{v}_s = \nabla \alpha$; $S$ is the entropy density; $N$ is the number of $\text{He}^3$ atoms per unit volume; the fluid unit volume momentum in the reference frame moving with the velocity $\mathbf{v}_s$ is expressed via variables $\beta$ and $\xi$ as

$$\mathbf{j} = \rho_n (\mathbf{v}_n - \mathbf{v}_s) = S \nabla \beta + N \nabla \xi$$

where $\rho_n$, $\mathbf{v}_n$ are the normal flow density and velocity. The mass flux density **I** equals to

$$\mathbf{I} = \rho_s \mathbf{v}_s + \rho_n \mathbf{v}_n = \mathbf{j} + \rho \mathbf{v}_s,$$

$\rho_s$ is the superfluid part density.

For investigating the nonlinear influence of optical waves on second sound-wave propagation we may limit ourselves with linear hydrodynamic equations [5] with a nonlinear optical source in the entropy equation

$$\frac{\partial \rho}{\partial t} = -\Delta(S_0 \beta + N_0 \xi + \rho_0 \alpha);$$

$$\frac{\partial \alpha}{\partial t} = -\mu; \quad \frac{\partial \beta}{\partial t} = -T; \quad \frac{\partial \xi}{\partial t} = -\zeta;$$

$$\frac{\partial S}{\partial t} = -S_0 \Delta \left( \frac{S_0}{\rho_{n0}} \beta + \frac{N_0}{\rho_{n0}} \xi + \alpha \right) + \frac{Q}{T};$$

$$\frac{\partial N}{\partial t} = -N_0 \Delta \left( \frac{S_0}{\rho_{n0}} \beta + \frac{N_0}{\rho_{n0}} \xi + \alpha \right). \qquad (1)$$

In these equations $\mu$ is the chemical potential of the solution, $\xi$ is the chemical potential of $He^3$ particles, $T$ is the temperature, $Q$ is the amount of heat emitted per unit volume and time unit due to light waves absorption. This quantity is proportional to the square modulus of the optical field $|E|^2$. The resonant for the second sound part of this quantity at the difference frequency is as follows

$$Q \sim \frac{c_l n \gamma_l}{8\pi} E_1 E_2^* \, e^{-i\Omega t},$$

where $c_l$ is the light velocity, $n$ – is the refraction index, $\gamma_l$ is the light waves amplitude damping coefficient, $\Omega = \omega_1 - \omega_2$, where $\omega_1, \omega_2$ are the frequencies of the light waves $E_1$ and $E_2$. Let the light waves with wave vectors $\mathbf{k}_1$ and $\mathbf{k}_2$ propagate at an angle $\vartheta$ to each other. In this case the wave vector $\mathbf{q}$ of the excited second-sound wave is

$$|\mathbf{q}| \approx 2\frac{\omega}{c} \sin\frac{\vartheta}{2},$$

here $\omega \approx \omega_1 \approx \omega_2$.

We shall write Eqs. (1) in terms of canonical variables. For this it is convenient to introduce new variables $v, \psi, \varphi$ and $\eta$ [5]: $\delta \mathbf{S} = \mathbf{S}_0 v$; $\psi = \mathbf{S}_0 \delta \boldsymbol{\beta}$; $\varphi = \rho_0 \delta \alpha$; $\delta \rho = \rho_0 \eta$. Here the following vector notations are used: $\mathbf{S} = (S, N)$, $\boldsymbol{\beta} = (\beta, \xi)$. The pairs ($v, \psi$) и ($\eta, \varphi$) are also canonically conjugated variables. Using these variables we obtain Eqs. (1) in the form (we omit '0' for undisturbed thermodynamic quantities):

$$\frac{\partial \eta}{\partial t} = -\frac{1}{\rho}\Delta(\varphi + \psi);$$

$$\frac{\partial v}{\partial t} = -\frac{1}{\rho}\Delta\left[\left(1 + \frac{\rho_s}{\rho_n}\right)\psi + \varphi\right] + \frac{1}{S+N}\frac{Q}{T};$$

$$\frac{\partial \varphi}{\partial t} = -\rho^2 \frac{\partial \mu}{\partial \rho}\eta - \rho\left(S\frac{\partial \mu}{\partial S} + N\frac{\partial \mu}{\partial N}\right)v;$$

$$\frac{\partial \psi}{\partial t} = -\rho\left(S\frac{\partial T}{\partial \rho} + N\frac{\partial \zeta}{\partial \rho}\right)\eta - \left[S\left(S\frac{\partial T}{\partial S} + N\frac{\partial \zeta}{\partial S}\right) + N\left(S\frac{\partial T}{\partial N} + N\frac{\partial \zeta}{\partial N}\right)\right]v. \qquad (2)$$

The Fourier-components of these variables for weak superfluid solutions $He^3 - He^4$ are expressed via second-sound normal coordinates $b_q$ as

$$\eta_q = \gamma B(b_q + b_{-q}^*),$$
$$v_q = (\gamma-1)B(b_q + b_{-q}^*),$$
$$iq\varphi_q = \Gamma^{-1}\rho c_2 B(b_q - b_{-q}^*),$$
$$iq\psi_q = -\Gamma^{-1}\rho c_2 B(b_q - b_{-q}^*), \quad (3)$$
$$\gamma = -\frac{\sigma}{\rho}\left(\frac{\partial\rho}{\partial\sigma}\right)_{c,p} - \frac{c}{\rho}\left(\frac{\partial\rho}{\partial c}\right)_{\sigma,p},$$
$$B = \left(\frac{\rho_s}{\rho_n}\frac{\Omega}{2\rho c_2^2}\right)^{1/2},$$

$\sigma$ being the entropy per unit mass, $c$ is the concentration, $p$ is the pressure.

Eliminating from Eqs. (2) first $\varphi$, $\psi$, then the quantity $\eta$, taking into account the relations (3) and the fact that there is a nonlinear source in the equation for entropy, we come to the following equation for the slow variation of the second-sound amplitude $v_q$ with distance:

$$\frac{dv_q}{dx} = \frac{\left[1 - c_2^{-2}\left(G + \rho\frac{\partial\mu}{\partial\rho}\right)\right]\frac{c_2 c_1 n}{T(S+N)8\pi}E_1 E_2^*}{\frac{1}{c_2^2}\frac{\rho_s}{\rho_n}\left(DG - \frac{\partial\mu}{\partial\rho}F\right) + \left(1 + \frac{\rho_s}{\rho_n}\right)\left(-\gamma G + \frac{F}{\rho}\right) - \gamma\rho\frac{\partial\mu}{\partial\rho} + D}. \quad (4)$$

In Eq. (4) the following notations are used:

$$D = S\frac{\partial\mu}{\partial S} + N\frac{\partial\mu}{\partial N};$$
$$G = S\frac{\partial T}{\partial\rho} + N\frac{\partial\zeta}{\partial\rho};$$
$$F = S\left(S\frac{\partial T}{\partial S} + N\frac{\partial\zeta}{\partial S}\right) + N\left(S\frac{\partial T}{\partial N} + N\frac{\partial\zeta}{\partial N}\right).$$

Let us make a numerical estimate of the distance $l$ at which the second-sound intensity is amplified from fluctuation values up to observable values. From Eq. (4) we have

$$v_q(l) - v_q(0)_{fl} \approx v_q(l) = Al,$$

where $A$ is the right-hand side of Eq. (4). For weak solutions the variable $v_q$ is expressed via second-sound intensity as

$$v_q = \left(\frac{I}{2\rho c_2}\frac{\rho_s}{\rho_n}\right)^{1/2}\frac{1}{c_2},$$

and for the interaction length $l$ we obtain

$$l = \left(\frac{I}{2\rho c_2} \frac{\rho_s}{\rho_n}\right)^{1/2} \frac{1}{c_2 A}.$$

It can be shown that the thermodynamic parameters entering the expression for $A$ are of the following order of magnitude (taking into account in particular that the first-sound velocity $c_1$ is essentially higher than that of second sound)

$$\chi = \frac{c}{\rho}\frac{\partial \rho}{\partial c} \ll 1; \qquad \gamma \sim \left(\frac{c_2}{c_1}\right)^2 \frac{\rho_n}{\rho_s} \ll 1 \text{ if } T \text{ is not close to } \lambda\text{-point;}$$

$$G + \rho\frac{\partial \mu}{\partial \rho} \sim = \chi c_1^2; \qquad DG \approx \left(N\frac{\partial \mu}{\partial N}\right)^2 \sim \chi^2 c_1^4; \qquad F\frac{\partial \mu}{\partial \rho} \sim \frac{\rho_n}{\rho_s} c_1^2 c_2^2;$$

$$G \sim \chi c_1^2; \qquad \frac{1}{\rho}F \sim \chi^2 c_1^2; \qquad \rho\frac{\partial \mu}{\partial \rho} \sim \frac{\rho_n}{\rho_s} c_2^2; \qquad D \sim \chi c_1^2.$$

Taking into account the above relations we can represent an order-of-magnitude value of the second-sound excitation length as

$$l \sim \left(\frac{I}{2\rho c_2} \frac{\rho_s}{\rho_n}\right)^{1/2} \chi \frac{\rho_s}{\rho_n}\left(\frac{c_1}{c_2}\right)^2 \frac{T(S+N)}{\gamma_l I_l},$$

where $I_l \sim \frac{|E|^2}{8\pi} c_l n$ is the light-wave intensity, $E \approx E_1 \approx E_2$.

To estimate numerically the distance $l$ we will use experimental data given in Refs. [6,7]:

$T = 1.5°K;\quad c_1 = 2.3\cdot 10^4\, cm/s;\quad c_2 = 3\cdot 10^3\, cm/s;\quad \frac{\rho_s}{\rho_n} \approx 3;\quad c < 0.1;\quad \chi \approx 0.02;$

$\rho = 0.14\, g/cm^3;\ S = \rho\sigma \approx 5\cdot 10^{21}\, cm^{-3};\ N = \rho c/m_3 \approx 10^{21}\, cm^{-3}$ ($m_3$ is the mass of the $He^3$ atom); $\omega \approx 3\cdot 10^{15}\, s^{-1};\quad \gamma_l \approx (10^{-2} - 10^{-3})\, cm^{-1}.$

For the light intensity $I_l \sim 10^3\, W/cm^2$ the second sound can be amplified up to intensity $\sim 10^{-3}\, W/cm^2$ at a distance $\sim 1\, cm$. Note, that from the point of view of an experiment in a superfluid orders-of-magnitude higher light-wave intensities are possible as well if using a pulse mode of light propagation (see, for ex. [8,9]). In this case the excitation of second sound could be considerably more effective.